\documentclass[12pt]{article} \usepackage{latexsym}
\usepackage{amsmath}

\begin{document}

\bibliographystyle{plain}

\def\one{{\bf 1}}
\def\dal{{\Box}}
\def\eps{{\epsilon}}
\def\cio{{\cal C}^\infty_0}
\def\ci{{\cal C}^\infty}
\def\Dp{{{\cal D}^{\prime}}}
\def\Sp{{{\cal S}^{\prime}}}
\def\Ep{{{\cal E}^{\prime}}}
\def\zero{{\bf 0}}
\def\k{{\bf k}}
\def\l{{\bf l}}
\def\wk{\omega_\k}
\def\lk{\lambda_\k}
\def\p{{\bf p}}
\def\bfsig{{\bf \sigma}}
\def\u{{\bf u}}
\def\x{{\bf x}}
\def\y{{\bf y}}
\def\z{{\bf z}}
\def\K{{\bf K}}
\def\R{{\bf R}}
\def\E{{\rm e}}
\def\ss{\mbox{ sing supp }}
\def\WF{\mbox{WF }}
\def\supp{\mbox{supp }}
\def\conesupp{\mbox{cone supp }}
\def\Srd{S_{\rho,\delta}}
\def\Lrd{L_{\rho,\delta}}
\def\Ird{I_{\rho,\delta}}
\def\half{{\frac{1}{2}}}
\def\GL{{\rm{GL}}}
\def\diag{{\rm diag}}
\def\real{{\rm Re}}



\def\slashi#1{\rlap{\sl/}#1}
%
\def\slashii#1{\setbox0=\hbox{$#1$}             
   \dimen0=\wd0                                 
   \setbox1=\hbox{\sl/} \dimen1=\wd1            
   \ifdim\dimen0>\dimen1                        
      \rlap{\hbox to \dimen0{\hfil\sl/\hfil}}   
      #1                                        
   \else                                        
      \rlap{\hbox to \dimen1{\hfil$#1$\hfil}}   
      \hbox{\sl/}                               
   \fi}                                         %
%
\def\slashiii#1{\setbox0=\hbox{$#1$}#1\hskip-\wd0\hbox
to\wd0{\hss\sl/\/\hss}}
%
\def\slashiv#1{#1\llap{\sl/}}

\newcommand{\fig}[1]{Fig.~\ref{fig:#1}}
\newcommand{\figlabel}[1]{\label{fig:#1}}

  \title{Stable, Renormalizable, Scalar Tachyonic Quantum Field Theory 
with Chronology Protection} 
  \author{Marek J. Radzikowski$^*$ ({\tt{radzik@physics.ubc.ca}}) \\
	\small{$^*$Present address: UBC Physics and Astronomy Dept.} \\ 
	\small{6224 Agriculture Road} \\
	\small{Vancouver, BC V6T 1Z1, Canada}}
  \date{\today}
  \maketitle

\baselineskip 2em    

\begin{abstract}
We use microlocal arguments to suggest that Lorentz symmetry breaking 
must occur in a reasonably behaved tachyonic quantum field theory 
that permits renormalizability. In view of this, 
we present a scalar tachyonic quantum field model with manifestly broken 
Lorentz symmetry and without exponentially growing/decaying modes. A notion of 
causality, in which anti-telephones are excluded, and which is viewed as
a form of chronology protection, is obeyed. The field theory is 
constructed in a preferred tachyon frame in terms of commuting 
creation/annihilation operators. We calculate some sample 
(renormalized) operators in this preferred frame, argue that the Hadamard
condition is satisfied, and discuss the PCT and spin-statistics theorems 
for this model.  
\end{abstract}

\section{Introduction}
Ever since the notion of {\it tachyons}, i.e., particles which always travel 
faster than light, was conceived as a possibility within the basic framework
of special relativity \cite{BilDesSud62}, a quantum field theory describing 
such 
particles has been sought. Despite numerous attempts to formulate such a theory
\cite{Tan60,Fei67,AroSud68,DhaSud68,Sch71}, it appears that a model which is 
consistent with the desiderata of conventional quantum field theory has been 
wanting \cite{KamKam71,KamKam78,HugSte90}. However, due to the interest in the
possibility that the neutrino may turn out to be a tachyon \cite{CHK85}, which
has been supported, at least na\"ively, by tritium beta decay experiments 
during the 90's \cite{PDG96}, one perhaps should be motivated to clarify
the issue either by presenting a viable model, or by showing that no viable 
tachyonic quantum field model, according to some reasonable definition of 
``viable'', can exist.

It is apparent from the early literature cited above, that various authors have
attempted to hold tightly to the assumption of {\it Poincar\'e invariance}, no 
matter what the consequences for the theory. The usual way of
requiring such invariance in the context of a scalar quantum field theory is 
by demanding that the {\it two-point distribution} $\left<0\left|\phi(x)\phi(y)\right|0\right>$
be left unchanged in value when the spacetime points $x,y$ are simultaneously replaced by 
$\Lambda x+a,\Lambda y+a$ respectively. (Here, $\left|0\right>$ is some notion of ``vacuum''
or ``ground'' state, $\phi(x)$ is an operator-valued distribution called the {\it field 
operator}, $\Lambda$ is a proper, 
orthochronous Lorentz transformation, and $a$ is a constant spacetime vector.)
However, such a restriction, under further suitable physical assumptions, as will be 
explained in Section \ref{SymmBreak}, must necessarily lead to a 
non-renormalizable theory, i.e., one which could not be incorporated into 
any renormalizable interacting (or self-interacting) theory, and for which the 
renormalized stress-energy tensor of the free field 
would make no sense. The argument used in Section \ref{SymmBreak} comes from 
the arena of {\it microlocal quantum field theory} \cite{Rad92,Rad96a}, 
which has proven to be reasonably successful, for example, in clarifying renormalization
in {\it quantum field theory on curved spacetime} \cite{BruFre00,HolWal02}.

From the standpoint of quantum field theory on curved spacetime, it is clear that,
applying the usual methods to construct a scalar model of tachyons on flat {\it
Minkowski} spacetime, one
obtains a quantum field theory satisfying the usual axioms of {\it Wightman positivity}
(namely the condition on the two-point distribution corresponding to the positive
definiteness of the Hilbert space inner product),
local commutativity, and the {\it Hadamard condition}, with, however, no guarantee
of {\it Poincar\'e invariance}. (Note that this does not nullify the existence
of Poincar\'e {\it symmetry}, which is always a symmetry of the underlying Minkowski
spacetime, but that one allows for the possibility that the construction of the various
quantities needed in the quantum field theory may manifest {\it spontaneous Poincar\'e
symmetry breaking}.) That one is permitted to effect this construction, 
follows from the quite general existence (and uniqueness) theorems for the typical 
quantities needed
for a quantum field theory, such as the advanced and retarded fundamental solutions
$\Delta_A,\Delta_R$ to the inhomogeneous wave equation on a globally hyperbolic 
spacetime \cite{Lic61,Ler63}. Indeed,
there is also an existence theorem (and, up to smooth function, a uniqueness theorem) 
for a Feynman propagator satisfying the Hadamard condition on such a spacetime 
\cite{DuiHoer72,Rad92,Rad96a}. Furthermore, the smooth part of the Feynman propagator may 
be chosen so that the corresponding two-point distribution satisfies Wightman 
positivity \cite{DuiHoer72}. Note that it would be at the point of introducing this
smooth function that the Poincar\'e symmetry may need to be broken.

Carrying out this construction for the {\it tachyonic} Klein-Gordon 
equation (i.e., with the opposite sign in the mass squared term) leads, however, to 
a theory which yields exponentially growing renormalized expectation values of 
typical observables, since in order to ensure that the strict local commutativity
condition is satisfied, the mode solutions of the wave equation which are 
exponentially growing and decaying in time must be included in the mode expansion 
of the field operator. This situation is unpalatable from a physical point of view,
since we would hope that a valid quantum field theory describing tachyonic 
neutrinos would at least preserve {\it stability}. To support this, one may
simply observe that neutrinos from SN1987A, which were detected after travelling 
a distance of 150,000 light years at very close
to the speed of light, have evidently manifested this property. A further aspect 
of the theory which may at first glance appear slightly disconcerting is that  
the two-point distribution does indeed manifest explicit breaking of Lorentz 
invariance {\it and} time 
translation invariance (but space translation invariance is maintained).

There is at our disposal, however, the simple procedure of deleting the 
undesirable exponentially growing/decaying modes from the theory. Since these
modes contribute only a smooth function to the two-point distribution, the resulting
theory still satisfies the Hadamard condition, and so retains the possibility of
being incorporated into a renormalizable interacting or self-interacting quantum 
field theory according to the criteria of Weinberg's theorem \cite{BruFre00}. The theory also 
maintains Wightman positivity since it is explicitly obtained from a mode sum.
Two new features (from the point of view of conventional QFT) remain or emerge (from the
construction described in the previous paragraph): 
the theory still breaks Lorentz invariance (but now preserves
both time and space translation invariance), and the usual notion of local 
commutativity (commuting fields at spacelike separation) is not satisfied. 

However, neither of these two features appears insurmountable. Indeed, the notion
of a preferred {\it tachyon frame}, ``seen'' only by the tachyons in the theory, 
would constitute one of the predictions of the theory. (Note particularly that 
the preferred tachyon frame is not one in which the speed of light takes on 
a special value $c$, different from what would be measured in other frames; rather 
$c$ is the same in every inertial frame, as in the usual formulation of special
relativity.) Furthermore, the local commutativity axiom is violated so weakly 
that physical signals, made up of tachyons described by this QFT, can be sent backward 
in time, but only 
to points which are spacelike separated from the sender. This is true even for devices which use
relays to attempt to send messages backward in time to the sender (called {\it 
anti-telephones}). Such devices cannot be constructed, according to this QFT, 
since the (tachyonic) particles required for such a device cannot be simultaneously 
created from any of the vacuum states allowed in the theory. We consider this 
property to be a manifestation of {\it chronology protection}, which has otherwise
appeared in the quite different context of QFT on curved spacetime \cite{Haw92,KayRadWal97}.    

Working in the preferred tachyon frame, we begin the construction of the scalar 
model from scratch, presenting some of the Green's functions, in Section \ref{ScaModel}.
Continuing in the preferred frame, we then utilize a Lagrangian approach to 
determine the (renormalized) Hamiltonian and 
momentum operators for a Hermitian scalar tachyon, as well as the charge operator 
for a charged scalar tachyon, in Section \ref{Ops}. 
Further elaboration on the Hadamard and chronology 
protection properties, as well as some remarks on the $PCT$ and 
spin-statistics theorems for this
particular model are given in a final Discussion section (Section \ref{Disc}).

Note that a different approach (in appearance) is adopted by \cite{CibRem96}, who 
also obtain a stable, causal QFT based on a preferred frame and a non-standard 
synchronization scheme, without, however, a
discussion of the renormalizability of their theory. Also, they obtain some first
results for the beta decay spectrum near the endpoint (for tachyonic neutrinos), 
as well as suggest an alternative 
mechanism to neutrino oscillations involving 3-body (tachyonic) decay channels. 
We conjecture that their QFT approach and ours can be mapped to each other 
by a suitable (general linear) coordinate transformation. It is hoped that, if these
two approaches are indeed found to be compatible, that the present formulation in terms
of the more familiar Minkowski coordinates would bring further clarity to the approach of
\cite{CibRem96}. 
   
\section{Necessity of Lorentz symmetry breaking}\label{SymmBreak}

Here we draw upon tools from the microlocal approach 
to quantum field theory. This approach grew out of the necessity of dealing 
with the singularities inherent in quantum field theory on curved spacetime \cite{BirDav82,
Ful89,Wal94} in a general and coherent manner. Specifically, one applies techniques and 
theorems from microlocal analysis \cite{Hoer71,DuiHoer72} to the Green's functions (or, rather,
distributions)
of the quantum field theory. Such distributions include the advanced and retarded fundamental 
solutions $\Delta_A,\Delta_R$ to the inhomogeneous wave equation, the Feynman propagator
$\Delta_F$ (and its complex conjugate), the two-point distribution $\Delta^{(+)}(x_1,x_2)=
\left<0\left|\phi(x_1)\phi(x_2)\right|0\right>$,
and, more generally, the $n$-point or Wightman distributions, which are the vacuum expectation
values of the $n$-fold products of the field operator $\phi(x)$ at the $n$ points
$x_1,\ldots,x_n$. We remark that this incorporation of microlocal techniques into 
quantum field theory on curved spacetime has led to, among other results, a characterization 
of the {\it global Hadamard condition} \cite{KayWal91} in terms of a restriction on the 
{\it wave front set} of the 
two-point distribution \cite{Rad96a}, a resolution of Kay's singularity conjecture 
\cite{Rad92,Rad96b}, and the development of an Epstein-Glaser-like approach to renormalization
on curved spacetime \cite{BruFre00,HolWal02}. Note that, as input to the renormalization 
programme, the two-point distribution must be kept globally Hadamard. Since the characterization
of this condition in terms of wave front sets is important in the present context, we start 
with an introduction to the main concepts involved in this characterization in the following paragraph.

Recall that the {\it singular support} of a distribution $F(x)$ consists of all the 
points $x$ at which $F$ is not smooth. (In order for $F$ to be smooth at
$x$, an open neighbourhood of this point must exist, on which $F$ and all its derivatives
exist and are finite.) The {\it wave front set} of $F$ is an extension of the 
singular support of $F$ consisting of the pairs $(x,k)$ where $x$ is in the singular support, 
and $k\ne 0$ is a direction in the cotangent space of the spacetime at $x$. The wave front set is 
{\it conic} in the sense that if $k$ is the second component of a point in the wave front
set, then so is $\alpha k$, where $\alpha>0$ is a real number. The directions $k$ indicate,
heuristically speaking, the directions in the ``local Fourier space'' at $x$ in which 
the ``local Fourier transform'' of $F$ near $x$, along with all its derivatives, does {\it not} 
decay more rapidly than any polynomial in the Euclidean distance as this distance tends 
to infinity. The wave front set is a general enough
construct that it can be defined on a curved spacetime as readily as on a flat spacetime, and,
on flat spacetime, it is defined for distributions more general than the tempered distributions (whose 
Fourier transforms exist). Furthermore, some general results for performing operations with 
such distributions on 
manifolds (such as multiplication, convolution, and restriction to a submanifold) have simple 
statements in terms of their wave front sets. For more precise
definitions, please see \cite{Hoer90}.\footnote{\label{fn1}Note that we have implicitly used a definition
of Fourier transform that differs from that of \cite{Hoer90} by an extra minus sign in the 
argument of the exponential function. This accounts for the apparent discrepancy in signs
in some formulae, e.g., the exponent in our formula Eq.(\ref{tpd}) is minus that in the formula
for a Fourier integral operator used in Theorem {8.1.9} of \cite{Hoer90}.}

We now describe the wave front set of a two-point distribution $F^{(+)}(x_1,x_2)$ 
(on a globally hyperbolic curved spacetime $(M,g)$) 
satisfying the Hadamard condition \cite{Rad92,Rad96a}.\footnote{For corrections to the
original proofs given in these references (which contain gaps), please consult
\cite{Koeh95a,SahVer01}.} The quadruple $(x_1,k_1,x_2,k_2)$ is in 
the wave front set precisely when $x_1$ and $x_2$ either coincide or are on the same null
geodesic. If $x_1=x_2$, the covectors $k_1,k_2$ are both null and $k_1=-k_2$. If $x_1\ne x_2$ 
(but $x_1$ and $x_2$ are still on the same null geodesic) then $k_1,k_2$ are null and tangent
to this null geodesic at the points $x_1,x_2$ respectively. Furthermore, $k_1$ is minus the 
parallel transport of $k_2$ from $x_2$ to $x_1$ along this null geodesic. An additional 
restriction is that $k_1$ is always pointing in the {\it future} time direction (for both 
$x_1=x_2$ and $x_1\ne x_2$). Thus, $k_2$ is 
always pointing in the {\it past} time direction. There are no other restrictions on the covectors;
hence all covectors satisfying the above criteria are included in the wave front set. If, in addition,
the spacetime is Minkowski, and translation invariance holds, the two-point distribution is a 
distribution of the difference variable $x=x_1-x_2$, and the wave front set of the distribution
in one variable $f(x)$ consists of pairs of points $(x,k)$ where $x$ and $k\ne 0$ are null, $k$ 
(as a vector) is parallel (or anti-parallel) to $x$, if $x\ne 0$, and $k$ is future pointing.  

We sketch the steps of the proof that for a tempered, Poincar\'e invariant two-point 
distribution satisfying the tachyonic Klein-Gordon equation $(\dal -m^2)\phi=0$, the Hadamard
condition cannot be satisfied. (Temperedness is assumed here in order to guarantee good behaviour
of renormalized observables, e.g., to avoid exponential growth of observables in time.) 
Translation invariance implies that we can write the two-point
distribution as a distribution of the difference variable $x=x_1-x_2$, i.e., as $u(x)$. Now
$\hat u (k)$ is also a tempered distribution, and the wave equation implies that $\hat u (k)$ 
is nonzero only for $k^2=-m^2$, which is a one-sheeted hyperboloid. Furthermore, Lorentz invariance 
of the distribution $u(x)$ implies
Lorentz invariance of $\hat u(k)$. The above considerations imply that $\hat u (k)$ may be written
as $c(k)\delta(k^2+m^2)$, where $c(k)$ is a smooth function on the one-sheeted hyperboloid. 
Since a spacelike $k$ may be mapped by a proper, orthochronous
Lorentz transformation to $-k$, we obtain $c(k)=c(-k)$ (in fact $c$ is a constant), and 
therefore $\hat u (-k)=\hat u (k)$. This implies $u(-x)=u(x)$, 
which, as is readily verified, leads to the result that $(x,k)\in\WF(u)$ if and only if 
$(-x,-k)\in\WF(u)$. Thus the wave front set covectors are not restricted to be future pointing;
past pointing ones are also required. (Note that the wave front set of the two-point
distribution must contain {\it some}
pairs, if this theory is to even vaguely resemble a typical nontrivial quantum field theory!) 
This finishes the sketch that the Hadamard condition 
cannot be satisfied for a Poincar\'e invariant, tempered two-point distribution of a tachyonic 
quantum field theory.

We note that a similar ``no-go'' result, in which the requirement of temperedness is dropped,
appears to have the following counterexample: First note that the symmetric part of the two 
point distribution (without
exponentially growing/decaying modes) is Poincar\'e invariant, while the same is true of the
commutator distribution (determined through the Leray-Lichnerowicz uniqueness theorem, which 
{\it requires} the presence of the growing and decaying modes). Thus, we may combine these terms to form
a ``hybrid'' two-point distribution which satisfies the Hadamard condition. However, Wightman
positivity evidently fails for this two-point distribution (at least it apparently cannot be constructed
from a mode sum). Hence it seems reasonable to conjecture that a Poincar\'e invariant two-point 
distribution, satisfying Wightman positivity and the tachyonic Klein-Gordon equation, {\it cannot}
satisfy the Hadamard condition. The author is aware only of arguments in favour of this conjecture
being true \cite{Fre98}. In any case, the constructed examples lend support to this conjecture's 
validity.    

A further observation is that, with the insertion of the extra assumption of Wightman positivity
for the two-point distribution in our no-go ``lemma'' for the tempered case, we can assert more
strongly that the two-point distribution becomes {\it real} $u(x)^*=u(x)$, in addition to being 
even $u(-x)=u(x)$. This is so because Wightman positivity implies that $\hat u(k)$ is real and
positive-valued (in the appropriate sense of distribution theory), besides being even 
$\hat u (-k) = \hat u (k)$. Hence the antisymmetric 
(and imaginary) part of $u(x)$ is zero. This appears to go far astray from describing a QFT with which we are 
presently familiar. In any case, the Hadamard condition still fails (this is now readily seen 
in the fact that the
usual leading order Hadamard singularity must have a non-zero imaginary part). As before, the breakdown of
this important input to renormalization forces us to consider such a model as {\it unphysical}
for our purposes. 

Finally, we note that non-Hadamard two-point distributions render the existence of the renormalized 
stress-energy tensor problematic, since the conditions on the wave front set
under which one can construct the 
two-point distribution of the Wick-ordered polynomials (of which the stress-energy tensor is one)
are not satisfied. See \cite{BruFreKoeh96,Wal77,Wal78} for more discussion on the stress-energy 
tensor and the need for the Hadamard condition to be satisfied in order that the stress-energy 
tensor be defined.  

\section{Construction of the scalar model}\label{ScaModel}

Having argued that one should expect Poincar\'e symmetry to be broken in a reasonable 
quantum field model satisfying the tachyonic Klein-Gordon equation, we now present such a
model. We shall, in this and the next sections, perform the constructions in just the 
preferred frame. Note that values of the scalar Green's functions in a boosted frame are readily 
obtained from these by the obvious change of coordinates. E.g., if a Green's function in
the preferred frame is $G(x',y')$, then the same function in the frame obtained by the boost
$\Lambda\colon x'\to x$ is the pullback by the inverse Lorentz transformation, namely 
$G(\Lambda^{-1}x,\Lambda^{-1}y)$.  

The first part of the construction in any frame is a listing of all the {\it oscillatory} 
mode solutions of the tachyonic Klein-Gordon equation (using $\dal=\partial_t^2 - \nabla^2$)
\begin{equation}\label{eq:TKG}
	(\dal - m^2)u=0\;.
\end{equation} 
In our choice of preferred frame, one has a natural choice of {\it inner product} for 
such modes, namely 
\begin{equation}\label{eq:InnPro}
  (u,v) = i\int_{t=a} u^*(x) \overleftrightarrow{\partial_t} v(x)\,d^3\x
\end{equation}
where 
\begin{equation}
  u^*(x)\overleftrightarrow{\partial_t} v(x) = u^*(x)(\partial_t v(x)) 
  - (\partial_t{u^*(x)}) v(x)\;. 
\end{equation}
This inner product turns out to be independent of the choice of the constant $a$
(as will be evident for the orthonormal basis we shall construct). 
However, the inner product {\it does} in general depend on the choice of
space-like hypersurface over which the integral is defined. This is in 
constrast to the case of massive or light-like particles. 

We choose (in this reference frame) a basis for the space of {\it positive energy} or
{\it positive frequency} 
(oscillatory) solutions as follows.
We consider the solutions $u_\k(t,\x) = M_\k\E^{-i(\wk t-\k\cdot\x)}$, where 
$\wk = \sqrt{\k^2 - m^2}$ and $|\k|>m$. In order to obtain the orthonormal relations
  \begin{eqnarray}\label{eq:RESoln}
    (u_\k,u_\l) &=& \delta^{(3)}(\k-\l) \\
    (u^*_\k,u^*_\l) &=& -\delta^{(3)}(\k-\l) \\
    \label{eq:RESoln2}	
    (u_\k,u^*_\l) &=& (u^*_\k,u_\l) = 0\;,
  \end{eqnarray}
we choose the normalization factors to be $M_\k = ((2\pi)^3\cdot
2\wk)^{-\half}$, whence
\begin{equation}
  \label{eq:PosFreMod}
  u_\k(t,\x) = {\frac{1}{\sqrt{(2\pi)^3\cdot 2\wk}}}\E^{-i(\wk
  t-\k\cdot\x)}\;.
\end{equation}

For mode solutions with $|\k|=m$, which are also oscillatory, the frequency is $0$, 
and thus time derivatives of the modes give zero. Thus the modes here are of the form
\begin{equation}
  \label{eq:ZeroFre}
  v_\k(t,\x) = \E^{i\k\cdot\x}
\end{equation}
with 
\begin{equation}
  \label{eq:NormZerFreq}
  (v_\k,v_\l) = 0\;.
\end{equation}
Hence any normalization constant will do here, and we just leave the
$v_\k$ as they are. Note that inner products with the $u_\k, |\k|>m$ are zero.
It will turn out that incorporating these modes into the theory does not
ultimately affect the two-point distribution, since the integrations
involving only these ``zero modes'' are over a set of measure zero. Hence we may safely
ignore them in the further development of the theory. 

Note that the {\it negative energy} or {\it negative frequency} solutions are simply taken to be
the complex conjugates of the positive energy modes. Thus they are 
\begin{equation}
  \label{eq:NegFreMod}
  u^*_\k(t,\x) = {\frac{1}{\sqrt{(2\pi)^3\cdot 2\wk}}}\E^{i(\wk
  t-\k\cdot\x)}\;,
\end{equation}
where the parameter space labelled by $\k$ is restricted to $|\k|>m$ as above.

As stated in the Introduction, the exponentially growing and decaying modes, which 
would be labelled by $|\k|<m$, are omitted from the model in order to avoid 
exponential blow-up of renormalized observables, and thus they shall be ignored
henceforth in this paper. A model in which these are incorporated (as partly 
described in the Introduction) was given in \cite{Sch71}, and also was considered
in the context of quantum field theory on curved spacetime by \cite{Rad98b,Fre98}. 

In order to quantize Eq.(\ref{eq:TKG}) without the exponentially growing/decaying modes,
we seek a field operator $\phi(x)$ of the following form
\begin{equation}
  \phi(x) = \int_{|\k|>m} (a_\k u_\k(x) + a^\dag_\k u^*_\k(x))\,d^3\k \;,
\end{equation}
which, besides satisfying Eq.(\ref{eq:TKG}), also satisfies the 
{\it equal time commutation relations}, modified so as not to include
frequencies $\k$ for which $|\k|<m$:
\begin{eqnarray}
  \label{eq:ETCR}
  \left[\phi(t,\x),\phi(t,\y)\right] &=& 0 \\
  \left[\phi(t,\x),\partial_t\phi(t,\y)\right] &=& i\delta_m(\x-\y)\;.
\end{eqnarray}
Here, $\delta_m$ is a modification of the Dirac delta distribution obtained
by deleting modes $\E^{i\k\cdot\x}$ with frequencies $|\k|<m$ from the 
latter, namely,
\begin{equation}
   \delta_m(\x)=\frac{1}{(2\pi)^3}\int_{|\k|>m}\E^{i\k\cdot\x}\,d^3\k\;.
\end{equation} 	 
  
We find the following relations:
\begin{equation}
  \label{eq:Rel1}
  (u_\k,\phi)=a_\k, \ (u^*_\k,\phi) = -a^\dag_\k\;,
\end{equation}
and
\begin{equation}
  \label{eq:Rel2}
  \left[a_\k,\phi(x)\right] = u^*_\k(x)\;,
\end{equation}
where we have employed
\begin{equation}
  \int u^*_\k(t,\y)\delta_m(\x-\y)\,d^3\y = u^*_\k(t,\x)
\end{equation} 
for $|\k|>m$. These relations lead to the commutators
\begin{equation}
  \label{eq:Comm}
  \left[a_\k,a_\l\right] = \left[a^\dag_\k,a^\dag_\l\right] = 0, \
  \left[a_\k,a^\dag_\l\right]=\delta^{(3)}(\k-\l)\;. 
\end{equation}

\def\gs{\left|0\right>}   
\def\gsb{\left<0\right|}  

The {\it vacuum} or {\it ground state} $\gs$ associated with this particular choice of preferred tachyon frame
is then defined by $a_\k\gs = 0$. The {\it two-point distribution} is therefore
\begin{eqnarray}
  \Delta^{(+)}(x,y) &=& \gsb\phi(x)\phi(y)\gs \\
  &=& \int_{|\k|>m} u_\k(x) u^*_\k(y)\,d^3\k \\
  &=& \frac{1}{2}\Delta^{(1)}(x,y) + i\frac{1}{2}\Delta(x,y)\;.
\end{eqnarray}
Taking the anti-symmetric part of the two-point distribution, namely,
\begin{equation}
	\gsb[\phi(x),\phi(y)]\gs =i\Delta(x,y) = 2i{\rm Im}\gsb\phi(x)\phi(y)\gs\;,
\end{equation}
we obtain the commutator distribution 
\begin{equation}
  \Delta(x,y) = -{\frac{1}{(2\pi)^3}}\int_{|\k|>m}d^3\k\,\E^{i\k\cdot(\x-\y)}
  {\frac{\sin[\sqrt{\k^2 -m^2}(t-s)]}{\sqrt{\k^2 - m^2}}}\;. 
\end{equation}
This is seen to be the unique distributional bisolution of the Cauchy
problem    
\begin{eqnarray}
  \label{eq:CauPro}
  \Delta(t,\x,t,\y) &=& 0 \\
  \partial_t\Delta(t,\x,s,\y)|_{s=t} &=& -\delta_m(\x-\y)\;.
\end{eqnarray}
The symmetric part of the two-point distribution, namely
\begin{equation}
	\Delta^{(1)}(x,y)= \gsb\{\phi(x),\phi(y)\}\gs = 
	2{\rm Re}\gsb\phi(x)\phi(y)\gs\;, 
\end{equation}
is
\begin{equation}
  \Delta^{(1)}(x,y) = {\frac{1}{(2\pi)^3}}\int_{|\k|>m}d^3\k\,
  \E^{i\k\cdot(\x-\y)}
  {\frac{\cos[\sqrt{\k^2 -m^2}(t-s)]}{\sqrt{\k^2 - m^2}}} \;.
\end{equation}
Note that this is a Lorentz invariant distribution.

\section{Renormalized operators}\label{Ops}

Taking the (classical) Lagrangian density of our field theory to be 
\begin{equation}
   {\cal L} = \frac{1}{2}\left(\partial^\mu\phi\partial_\mu\phi + m^2\phi^2\right)\;,
\end{equation}
where the field is real-valued, we obtain, via the usual method, the Hamiltonian 
\begin{equation}
  H=\frac{1}{2}\int \left(\pi^2 + (\nabla\phi)^2 - m^2 \phi^2\right)\,d^3\x\;,
\end{equation}  
where $\pi(x) = \partial_t\phi(x)$. Note that, because of the minus sign 
in the third term above, the Hamiltonian is generally not positive. 
However, we shall shortly show that the quantized version is indeed a positive
operator in the preferred frame. Similarly, the momentum of the field is
\begin{equation}
  {\bf P} =-\int \partial_t\phi\nabla\phi\,d^3\x\;.
\end{equation}  

The quantization of the above expressions proceeds along the familiar lines:
classical fields are replaced by the quantized versions, and products of field 
operators are normal ordered. We express the final result in terms of 
creation/annihilation operators, recalling that normal ordering places creation 
operators before annihilation operators in the expressions. Thus, we obtain
\begin{eqnarray}
  \pi &=& \partial_t\phi = -i\int_{|\k|>m} \wk (a_\k u_\k - a^\dag_\k u^*_\k)\,d^3\k \\
  \nabla\phi &=& i\int_{|\k|>m} \k(a_\k u_\k - a^\dag_\k u^*_\k )\,d^3\k\;,
\end{eqnarray}
and, using the orthonormality property of the exponential functions $\E^{i\k\cdot\x}$, 
\begin{eqnarray}
  H &=& \int_{|\k|>m} \wk a^\dag_\k a_\k\,d^3\k \\
  {\bf P} & = & \int_{|\k|>m} \k a^\dag_\k a_\k\,d^3\k\;.
\end{eqnarray}
Note that these expressions are as expected, in analogy with the usual 
creation/annihilation operator formalism in conventional QFT. We anticipate that
these operators will transform in the usual manner when we calculate them in
a boosted frame. Hence the Hamiltonian, although it will remain Hermitian, 
will not remain positive in any frame 
boosted with respect to the preferred one.  

For the case of a charged scalar tachyon, the Lagrangian density becomes
\begin{equation}
   {\cal L} = \partial^\mu\phi^*\partial_\mu\phi + m^2\phi^*\phi\;,
\end{equation}
where the field is now complex-valued. The conserved quantity derived
from global phase invariance (i.e., invariance under $\phi \to \E^{i\alpha}\phi$) is
\begin{equation}
  Q = -iq\int\phi^*\overleftrightarrow{\partial_t} \phi\,d^3\x\;.
\end{equation}
The ansatz for the field operator is now
\begin{equation}
  \phi = \int_{|\k|>m} (a_\k u_\k + b^\dag_\k u^*_\k)\,d^3\k\;,
\end{equation}
while the quantized version of the above conserved quantity evaluates to
\begin{eqnarray}
  Q&=&-iq\int\colon\phi^\dag\overleftrightarrow{\partial_t}\phi\colon\,d^3\x	 \\
   &=& q\int_{|\k|>m}(b^\dag_\k b_\k - a^\dag_\k a_\k)\,d^3\k\;.
\end{eqnarray}
Thus the operator $b^\dag_\k$ may be interpreted as creating a particle of 
charge $q$ from the vacuum, while $a^\dag_\k$ may be interpreted as creating
an anti-particle of charge $-q$ from the vacuum. 

\section{Discussion}\label{Disc}

A sketch that the two-point distribution constructed in Section \ref{ScaModel}
is of the Hadamard form is now presented. One notes that the two-point 
distribution (of the difference variable) for our tachyonic theory may be written as
\begin{equation}\label{tpd}
  \Delta^{(+)}(x) = \frac{1}{(2\pi)^3}\int_{|\k|>m} \frac{1}{2\wk}\E^{-i(\wk t-\k\cdot\x)}\,d^3\k \;.
\end{equation} 
It is simple to show that the {\it scaling limit} of our two-point distribution is
\begin{equation}
  \lim_{\lambda\to 0}\lambda^2\Delta^{(+)}(\lambda x) = \Delta^{(+)}_0 (x)\;,
\end{equation}
which is the two-point distribution for the massless theory. Thus the leading order singularity
of $\Delta^{(+)}(x)$ is the usual Hadamard one. Conceivably, $\Delta^{(+)}(x)$ may have lower order 
singularities different from the ones emanating along the light cone from the origin, or the 
lower order singularities may produce extra covectors for the singularities emanating from the origin.
However, noting that Eq.(\ref{tpd}) describes a {\it Fourier integral operator} \cite{Hoer71}, 
with {\it homogeneous phase function} $\phi(x,\k)=|\k|t-\k\cdot\x$, we may avail ourselves of Theorem
8.1.9 of \cite{Hoer90},\footnote{See Footnote \ref{fn1}.} which restricts the pairs $(x,k)$ in the wave front set to be precisely
those in the leading order singularity. Thus the wave front set of $\Delta^{(+)}$ is of the Hadamard  
type. Since the anti-symmetric part of $\Delta^{(+)}$ is the same as $i$ times the
advanced minus retarded fundamental solutions, up to smooth function, the equivalence theorem of 
\cite{Rad96a} tells us that $\Delta^{(+)}$ satisfies the global Hadamard condition. Thus, we expect that
the free field theory presented in Section \ref{ScaModel} may be used as input to a renormalizable 
(self-) interacting theory, satisfying the criteria of Weinberg's theorem \cite{BruFre00}. That it leads
to sensible renormalized expectation values of observables quadratic in the fields has already been 
demonstrated for some typical cases in Section \ref{Ops}.   

We now explain the ``no anti-telephones'' property of this model, which we consider to be a version 
of {\it chronology protection}, as mentioned in the Introduction. An {\it anti-telephone} is a device
in which a relay at $B$ is set up at a spacelike separation to a message sender at $A$. The relay is designed
to receive a message sent via tachyons from $A$ and immediately to resend the message, again using tachyons, 
back to a receiver at $A'$ (the same individual as the initial sender, but at a different time). 
The idea of the anti-telephone
is to attempt to violate causality by sending the relayed message from $B$ to $A'$ {\it more backward in time} (resp. less forward in time) 
than the starting message was sent from $A$ to $B$ forward in time 
(resp. backward in time). Then one would have used tachyons to effect 
what should be considered
a highly egregious violation of causality. (If this could be done, then a signal could be sent whose effect would
be to prevent the message from being sent in the first place. However, if the message had not been sent, 
then nothing would have hindered it from being sent.) However, it is apparent that, due to the cutoff in the spectrum
of the one-particle states of the model presented in Section \ref{ScaModel}, there is a global directional
dependence in the lower bound of the allowed energies, such that, if the tachyons managed to travel backward in time
in moving from $A$ to $B$, they would certainly be forced to travel as much forward in time, or more so, in travelling
back from $B$ to $A'$. Only particles with such a directionally dependent lower bound in the energy 
may be created out of the vacuum either in the preferred frame, 
or in any other frame, boosted with respect to the preferred one. Similarly, no sequence of relays could be constructed
to guide the tachyons along some path so that they arrive back to $A'$ at a previous time to $A$, since it is clear that 
such a path of tachyonic world-lines, any one of which goes backward in time, is not constructible in the 
preferred frame. If it is impossible in one inertial 
frame, then it must be so for all inertial frames. Note that we should postulate that the same preferred frame 
must be universal 
to all tachyonic particles in the same interacting theory, since otherwise, severe violations       
of causality, in principle, could be brought about. 

It is of interest to determine whether any of the ``big theorems'' of axiomatic QFT  
\cite{StreWig64,Jos65} remain in our model, or in 
models constructed in a similar approach (by restricting $4$-momenta to lie in the upper half of a single-sheeted 
hyperboloid cut through by a spacelike hyperplane through the origin). As a first step, we consider the PCT theorem
for the Hermitian scalar model. The relevant property to be proved for the two-point distribution
is
\begin{equation}
  \Delta^{(+)}(x)^* = \Delta^{(+)}(-x)\;.
\end{equation}  
This is equivalent to the statement that the Fourier transform of $\Delta^{(+)}$ is real-valued, which 
is evidently true for our model in any inertial frame. (The Fourier transform of the two-point distribution is a positive 
multiple of $\theta(k^0+\beta k^z)\delta(k^2+m^2)$ in a frame boosted by a speed $\beta$ in the $z$ direction
relative to the preferred frame.)  Hence a PCT theorem holds for this model. (The above property, appropriately reformulated,
extends to all the Wightman distributions, since the theory we have constructed is {\it quasi-free}, and the
Wightman distributions determine the full theory by the Wightman reconstruction theorem \cite{Wig56}.) 
 
Next, we touch upon the spin-statistics theorem. We would expect that a well-behaved tachyonic model would
reduce to a physically well-behaved massless theory as the tachyonic parameter $m$ tends to $0$. That would mean that 
the G\aa rding-Wightman axioms (Wightman positivity, Poincar\'e invariance, spectral condition, local commutativity) 
should hold for the scaling limit two-point distribution. However, if the wrong 
connection between spin and statistics is assumed in this case, then one would necessarily obtain a non-Lorentz
invariant theory in the scaling limit, since all the other properties would presumably be satisfied for this
limit. (If all the axioms hold in the limit, the limit two-point distribution must be zero, by the usual spin-statistics
theorem. However, this would contradict the definition of the scaling limit as the leading order [non-zero] behaviour 
of the two-point distribution as the difference variable $x$ tends to $0$, unless, of course, the two-point distribution of the original tachyonic
theory is itself $0$.) Thus, to avoid this undesirable failure of Lorentz invariance in the scaling limit, we must 
retain the usual connection between spin and statistics. Note that this model then stands in constrast to the one suggested
by Feinberg \cite{Fei67}, who assumed the wrong connection of spin with statistics, e.g., anti-commutation relations
in the scalar theory.     

Finally, we observe that the spacelike hypersurface through the origin (in Fourier space), which bounds the upper
half of the single-sheeted mass hyperboloid from below (i.e., the one-particle spectrum of the model described
here), may be regarded as defining a {\it frame-dependent interpretation rule} for the allowed $4$-momenta of particles
and anti-particles in the QFT. This is, in effect, a use of the ``Re-interpretation Principle'' of \cite{BilDesSud62},
which proposes to regard a negative energy, backward-in-time-moving particle/anti-particle of momentum $\k$ as a positive energy, 
forward-in-time-moving anti-particle/particle of momentum $-\k$. This would at first seem to suggest an identification
of the $4$-momentum $k$ with $-k$ on the full single-sheeted hyperboloid. However, we find it more appropriate to pick
a {\it single} description from each pair $(k,-k)$, to describe {\it both} a particle and anti-particle, and to do so in each 
frame in a way that preserves chronology protection (a plane must be used to cut the hyperboloid) and the Hadamard condition
(the half containing arbitrarily large {\it positive} energies must be chosen), and is consistent with Lorentz covariance. 
(The fact that, in the quantum field model, we choose a single description from among two descriptions which are equally valid 
from the classical viewpoint, suggests an {\it economy of description} principle inherent in the quantum field theory of
tachyons. This, of course, is satisfied in the usual choice of the upper mass hyperboloid [out of two sheets], as is made in the 
regular massive Klein-Gordon theory.) Such a ``halving'' of the single sheeted mass hyperboloid is, up to a boost, unique.
We find it rather remarkable that such a simple interpretaion rule leads to {\it both} chronology protection {\it and} the Hadamard 
condition (i.e., renormalizability) being satisfied. This points to a deep unity among the ``axioms'' which we adopt as
physical.  

In conclusion, we expect that future work in this subject will be done to further develop and 
clarify the quantum field theoretic aspects of tachyons, especially those involving 
interactions. A basic step in this direction has been to clarify the calculation of the phase
space factor that appears in two-body decay in which one of the products is a tachyon \cite{Rad08a}.
Note that, in that paper, the underlying quantum field theory is implicitly assumed to be in 
accord with the model presented in this paper (in Section \ref{ScaModel}). We also foresee the development
started here as extending consistently to Dirac-like tachyonic ({\it Dirachyonic}) 
quantum field theory, whose ramifications (especially the inherent maximal
parity breaking that arises in such a model) would tend to support the 
possibility that the neutrino may be a tachyon. 

\vspace{1em}\noindent
{\bf Acknowledgements:} I wish to thank A.S. Wightman and K. Fredenhagen for first asking me questions about
tachyonic QFT which led (indirectly or directly) to the present attempt at formulating such a theory
consistently. Also, G. Heinzelmann, B.S. Kay, R. Atkins, A. Chodos, W.G. Unruh, and the UBC Dept. of Physics and 
Astronomy are acknowledged for helpful discussions, encouragement, and hospitality. 

\bibliography{big}
\end{document}